\documentclass[prb,aps,twocolumn,superscriptaddress,showpacs,floatfix]{revtex4}
\usepackage{graphicx}
\usepackage{amsmath}
\usepackage{bm}

\begin{document}

\title{Impurity effects at finite temperature \\ in the two-dimensional $S=1/2$
Heisenberg antiferromagnet}

\author{Kaj H. H\"oglund} 
\affiliation{Department of Physics, {\AA}bo Akademi University, 
Porthansgatan 3, FIN-20500, Turku, Finland}

\author{Anders W. Sandvik} 
\affiliation{Department of Physics, {\AA}bo Akademi University, 
Porthansgatan 3, FIN-20500, Turku, Finland}
\affiliation{Department of Physics, Boston University, 590 Commonwealth Avenue,
Boston, Massachusetts 02215, U.S.A.}

\date{\today}


\begin{abstract}
We discuss effects of various impurities on the magnetic susceptibility
and the specific heat of the quantum $S=1/2$ Heisenberg antiferromagnet on a
two-dimensional square lattice. For impurities with spin $S_i > 0$
(here $S_i=1/2$ in the case of a vacancy or an added spin, and $S_i=1$ for a
spin coupled ferromagnetically to its neighbors), our quantum Monte Carlo
simulations confirm a classical-like Curie susceptibility contribution
$S_i^2/3T$, which originates from an alignment of the impurity spin with the
local N\'eel order. In addition, we find a logarithmically divergent
contribution, which we attribute to fluctuations transverse to the
local N\'eel vector. We also study frustrated and nonfrustrated bond
impurities with $S_i=0$. For a simple intuitive picture of the impurity
problem, we discuss an effective few-spin model that can distinguish between
the different impurities and reproduces the leading-order simulation data
over a wide temperature range.
\end{abstract}

\maketitle

\section{introduction}\label{sec_intro}

The problem of impurities in low-dimensional quantum antiferromagnets has
attracted considerable attention ever since the discovery of high-temperature
superconductivity in the cuprates.\cite{manousakis} At low concentration,
holes doped into the CuO$_{\rm 2}$ planes are localized, or have very
low mobility, and hence static impurities are relevant for understanding the 
initial reduction of antiferromagnetism upon doping compounds such as 
La$_{\rm 2}$CuO$_{\rm 4}$.
These impurities are expected to be magnetically frustrated.\cite{aharony}
Although not directly related to the breakdown of antiferromagnetism
associated with the onset of superconductivity, static nonfrustrating
impurities, e.g., inert holes corresponding to substitution of Cu atoms by
nonmagnetic Zn,\cite{doping,greven} also can give important information
pertaining to the nature of the interactions in the CuO$_{\rm 2}$ planes.
The same applies to related cuprates where the planes are broken up
into chains \cite{chainexp} or ladders.\cite{ladderexp} Very recently,
similar impurity problems were also suggested to be of relevance to possible
physical realizations of quantum computers.\cite{castro_neto}

On the theoretical side, Heisenberg impurity models can be studied by a
wide range of modern quantum many-body methods. Importantly, numerical
techniques, such as quantum Monte Carlo and the density matrix
renormalization group, can give approximation-free results against which
analytical approaches can be tested on a quantitative level. Once such a
program has been completed, the applicability of a Heisenberg description
to an experimental system can be judged without concerns about approximations
in the calculations. There is already a large body of work devoted to various
impurity effects, and a coherent picture is emerging. Restricting ourselves
to work on single impurities, we note several ground state calculations for
Heisenberg chains,\cite{egg92a,iga95a,nis00a,nis00b} ladders,
\cite{fuk96a,san97a,mar97a} and the two-dimensional (2D) square lattice.
\cite{bul89a,nagaosa,sch94a,san97a,kot98a,nag95a,sus00a,sachdev1,sac01a}
Extensive work on chains \cite{egg95a,egg98a,fuj03a} and 2D systems
\cite{mur97a,sachdev1,paper1,sachdev2,sushkov} at finite temperature has
also been carried out. In this paper, we continue our studies \cite{paper1}
of finite-temperature effects of isolated static impurities in the
standard 2D Heisenberg model, and also present some results for the 3D system.

In a recent comprehensive quantum field-theoretical work,\cite{sachdev1} a
low-temperature theory of an arbitrary quantum impurity in a 2D
antiferromagnetic host system was developed, with the host being either
in the $T=0$ magnetically ordered phase (i.e., in the renormalized-classical
regime at $T > 0$) or close to a quantum-critical point. In the magnetically
ordered phase a leading-order classical-like Curie contribution to the
impurity susceptibility was predicted to stem from the coupling of the
impurity moment $\mathbf{S}_i$ to the local N\'eel order as the temperature
$T\to 0$, i.e., $\chi _{\text{imp}}^z \to S_i^2/3T$. Stimulated in part by
these theoretical predictions by Sachdev {\it et al.},\cite{sachdev1} we
recently carried out a large-scale quantum Monte Carlo (QMC) study
\cite{paper1} of the 2D $S=1/2$ Heisenberg antiferromagnet and confirmed the 
Curie prefactor $1/12$ in the renormalized-classical regime, for a
vacancy (missing spin) as well as for an added $S_i=1/2$ impurity spin.
However, we also discovered a low-$T$ logarithmically divergent subleading
contribution to the impurity susceptibility. This anomaly was attributed to
the transverse component, for which a
$T$-independent behavior had been predicted.\cite{sachdev1,sepnote1} Related
logarithmic divergences had also previously been found, e.g., in an exact
study of an impurity in the classical 2D Heisenberg model,\cite{harris} and
in a Green's function treatment of the 2D quantum model with an extra spin
at $T=0$.\cite{nagaosa} In the latter study, the frequency dependent
transverse impurity susceptibility $\chi_{\text{imp}}^{\perp}(T=0,\omega)$
was found to be log divergent when $\omega \to 0$. More recently, an
anomalous susceptibility was also found in the Heisenberg model with a
finite impurity concentration.\cite{chernyshev}

Recent efforts, by Sachdev and Vojta \cite{sachdev2} and Sushkov,\cite{sushkov}
to explain our previous numerical findings,\cite{paper1} have resulted in more
complete field-theoretical descriptions where the impurity susceptibility
indeed acquires a previously unnoticed subleading log divergent contribution.
Its principal cause can also in these analytical treatments be considered as
associated with the transverse component, although there are also higher-order
logarithmic contributions arising from longitudinal fluctuations.
\cite{sachdev2,sepnote1} In the formulation by Sachdev and Vojta,
\cite{sachdev2} which relies on an expansion of the nonlinear $\sigma$ model
around dimensionality $d=1$, a very detailed form for the logarithmic
corrections to the impurity susceptibility was given. For general impurity
spin $S_i$:
\begin{eqnarray}
\chi _{\text{imp}}^{z}&=&\frac{S_i^2}{3T}\left [1+\frac{T}{\pi
                        \rho _{s}}\ln \left (\frac{C_1\rho _s}{T}\right )
                        \right.
\nonumber \\
&&-\left.\frac{T^2}{2\pi ^2\rho _{s}^2}\ln \left (\frac{C_2\rho _s}{T}\right )
+O\left (\frac{T}{\rho _s}\right )^3\right ],
\label{subir}
\end{eqnarray}
where $\rho _s$ is the spin stiffness of the bulk-ordered antiferromagnet
in the absence of impurities and the unknown constants $C_{1,2}$ are in
general nonuniversal, but become universal when a quantum critical point is
approached. The first subleading term $\propto \ln (1/T)$ in Eq.~(\ref{subir})
hence accounts for the log divergent behavior observed in our numerical
studies. The quantitative agreement between Eq.~(\ref{subir}) and our
numerical data is quite remarkable, as will be shown in this paper. Our
results also agree qualitatively with the analytical results obtained
by Sushkov.\cite{sushkov}

The purpose of this paper is to give a more complete numerical account of the
effects of different types of single static impurities on the magnetic
susceptibility of the 2D $S=1/2$ Heisenberg antiferromagnet on a square
lattice. Some of the results were previously summarized in
Ref.~\onlinecite{paper1}. The impurity effects were there determined for a
vacancy and an added-spin impurity, by calculating impurity susceptibilities
with the stochastic series expansion (SSE) QMC technique.\cite{sse1,sse2} The
impurity susceptibility is simply the difference between the susceptibilities
of the pure and doped Heisenberg models. In this paper we compare our
numerical results for the vacancy and added-spin impurity models with the
theoretical expression in Eq.~(\ref{subir}). We also consider an impurity
consisting of a spin coupled ferromagnetically to its four nearest neighbors
[see Fig.~\ref{fig1}(d)]. This coupling arrangement is nonfrustrating and
can be expected to lead to an $S_i=1$ impurity, in contrast to the $S_i=1/2$
vacancy and added-spin impurities. It was suggested by Aharony
\textit{et al.},\cite{aharony} that hole doping the parent compounds of the
cuprate superconductors could lead to effective frustrated ferromagnetic
exchange couplings between nearest neighbor Cu spins. Motivated by this
scenario, we have also considered an impurity model with a single
ferromagnetic bond, and compared this with a missing bond. The 2D Heisenberg
antiferromagnet with two vacancies on different sublattices, and at different
separations, is also studied in order to further elucidate the behavior of
the single-vacancy impurity susceptibility. Finally, we have considered a
single vacancy in the three-dimensional (3D) Heisenberg antiferromagnet, for
which analytical limiting expressions has also been obtained recently.
\cite{sachdev2} Although the main focus of this paper is on the
susceptibility, we will also present some results for impurity effects on
the internal energy and the specific heat.

In Ref.~\onlinecite{paper1} we also introduced effective models for the
vacancy and added-spin systems. These models are very simple few-spin
systems constructed in order to capture the leading-order impurity
effects---they do not contain the log corrections. They provide simple
physical pictures of the dominant mechanisms at play in the full Heisenberg
impurity models. In this paper the effective models are  discussed in detail,
and the concept is further demonstrated by results for added-spin impurities
with different couplings to the host and the ferromagnetically coupled
in-plane impurity spin.

The rest of the paper is organized as follows. In Sec.~\ref{sec_models} the
full Heisenberg and effective models, as well as their impurity
susceptibilities, are defined. In Sec.~\ref{sec_method} the SSE Monte Carlo
method is briefly outlined, and the improved estimators needed to achieve
sufficient statistical accuracy are discussed. We also describe an
averaging trick used to alleviate the sign problem in our study of the
frustrated ferromagnetic-bond impurity. The SSE results are presented and
compared with the corresponding effective models in Sec.~\ref{sec_results}.
Concluding remarks are given in Sec.~\ref{sec_summary}. In the Appendix
we discuss the specific heat of the pure Heisenberg model, for which
we have obtained low-temperature results of unprecedented accuracy. In 
order to provide benchmark results for other calculations, we also list 
some selected high-precision numerical values for energies and 
susceptibilities of systems with different impurities.

\section{Impurity models and susceptibilities}\label{sec_models}

Following Ref.~\onlinecite{sachdev1}, an impurity susceptibility is defined
as the difference between the susceptibility of an impurity system and
the pure system, i.e.,
\begin{equation}
\chi _{\text{imp}}^{z,(i)}=\chi ^{z}_{(i)}-\chi^{z}_{(a)},
\label{chi_imp}
\end{equation}
where $i=b,c,d,e$, and $f$, correspond to the different impurity
systems shown in Figs.~\ref{fig1} and \ref{fig2}, and $\chi^{z}_{(a)}$ is
the susceptibility of the pure system. The susceptibilities on the right-hand
side of Eq.~(\ref{chi_imp}) are not normalized by the system size, i.e.,
\begin{equation}
\chi^{z}_{(i)}=\frac{1}{T}\left (\sum_j S^z_j\right )^2,
\label{chidef}
\end{equation}
where the sum is over all the spins of the pure or impurity systems. The
impurity susceptibilities are hence intensive differences of extensive
quantities, and they provide a natural framework for quantifying the effects
of different isolated impurities on the susceptibility of the pure system.
They also give the leading (linear) dependence on the concentration of
impurities. The definition in Eq.~(\ref{chi_imp}) will be used both in the
context of the full Heisenberg models and the corresponding effective models,
both of which will be defined in this section. Quantities analogous to
Eq.~(\ref{chi_imp}) will also be used for the internal energy and the
specific heat.

The impurity susceptibility can be separated in components parallel and
perpendicular to a given direction. Here the separation is done with respect
to an axis along the orientation of the local N\'eel order at the impurity.
In the isotropic 2D Heisenberg antiferromagnet, true long-range N\'eel order
sets in, i.e., the spin-rotation symmetry of an infinite system is broken,
only at $T=0$.\cite{chn} The components
$\chi _{\text{imp}}^{\parallel ,(i)}$ and $\chi _{\text{imp}}^{\perp ,(i)}$,
where $\parallel$ and $\perp$ refer to directions parallel and perpendicular
to the N\'eel order, are, therefore, true physical observables only at $T=0$.
However, our calculations show a temperature behavior that confirms an
approximate, but conceptually useful, 
separation of the impurity susceptibility
in components already at low finite $T$, as will be shown in
Sec.~\ref{sec_results}. In the 3D Heisenberg antiferromagnet, N\'eel order
is present already at finite temperature, below $T_{\rm c}/J\approx 0.95$,
\cite{sandvik3d} which makes the two components truly distinguishable. The
effective impurity models are defined to include a fluctuating direction
given by a classical vector $\mathbf{N}$, describing a local N\'eel order
with respect to which susceptibility components can be defined. Comparisons
with the QMC results show that the separation into components is useful even 
at a quantitative level.

\subsection{Full Heisenberg models}

\begin{figure}
\includegraphics[width=8cm,clip]{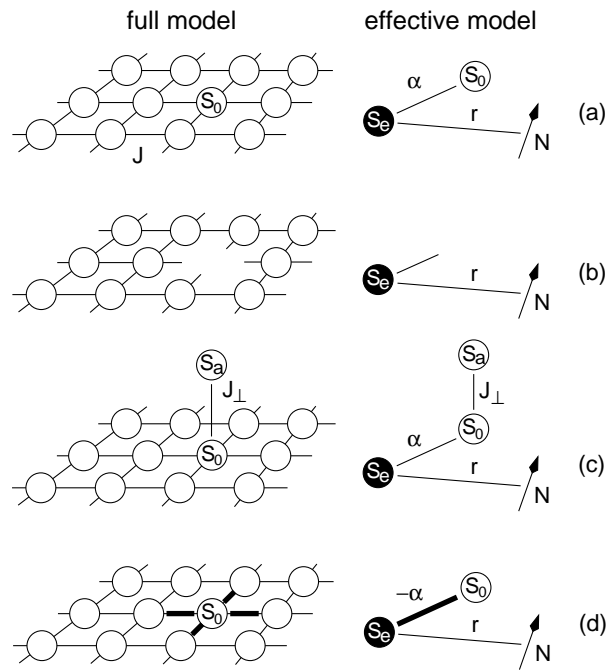}
\caption{Full Heisenberg models and corresponding effective models of the 
(a) pure, (b) vacancy, (c) added-spin, and (d) four ferromagnetic bonds 
systems. Thick solid lines symbolize ferromagnetic spin-spin couplings
$-J_F<0$, with $J_F=J$. The free parameters of the effective models are
the couplings $\alpha$ and $r$.}
\label{fig1}
\end{figure}

The basis for this study is the isotropic $S=1/2$ Heisenberg antiferromagnet 
on a periodic $L\times L$ lattice. This model is defined by the Hamiltonian
\begin{equation}
H_{(a)}=J\sum_{b=1}^{N_b}\mathbf{S}_{i(b)}\cdot \mathbf{S}_{j(b)},
\label{pure}
\end{equation}
where $J>0$, bond $b$ connects the nearest-neighbor sites $[i(b),j(b)]$,
and $N_b$ is the total number of bonds. $H_{(a)}$ is given a pictorial
representation to the left in Fig.~\ref{fig1}(a), and will hereafter
be referred to as the full Hamiltonian of the pure system. Impurity
models are obtained by introducing single defects in the pure model.

We begin by presenting the models with impurity moments $S_i\neq 0$. They
are illustrated in Fig.~\ref{fig1}. When a single spin $\mathbf{S}_0$
is removed from the square lattice, the vacancy model shown to the left in
Fig.~\ref{fig1}(b) is obtained. We will study it in 2D as well as in 3D.
The added-spin model, shown to the left in Fig.~\ref{fig1}(c), is
obtained by coupling a single off-plane spin-$\frac{1}{2}$ $\mathbf{S}_a$
antiferromagnetically to a spin $\mathbf{S}_0$ on the square lattice. Two
different values on the coupling strength $J_{\perp}=J$ and $J_{\perp}=J/2$
will be considered here. In the limit $J_\perp \to \infty$, the magnetic
properties of the added-spin model become equivalent to the vacancy model, 
since the two spins $\mathbf{S}_a$ and $\mathbf{S}_0$ are then locked in a 
singlet state. An $S_i=1$ impurity is obtained by considering a configuration
of four ferromagnetic bonds with one spin in common, as shown in 
Fig.~\ref{fig1}(d).

\begin{figure}
\includegraphics[width=5cm,clip]{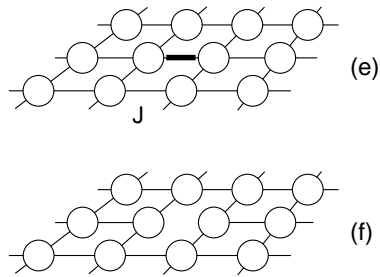}
\caption{Full Heisenberg models of the (e) frustrating ferromagnetic bond
($J_F=J$) and (f) removed bond ($J_F=0$) systems.}
\label{fig2}
\end{figure}

We also consider the models with impurity moments $S_i=0$ shown in
Fig.~\ref{fig2}. A system with one frustrating ferromagnetic bond, where
the spin-spin coupling is $-J_F<0$, is shown in Fig.~\ref{fig2}(e). In the
limit $J_F/J\to \infty$, one might hence expect a corresponding $S_i=1$
impurity moment, as the two spins connected by the ferromagnetic bond then
form a triplet. On the other hand, for $J_F=0$ clearly $S_i=0$, and hence a
transition between $S_i=0$ and $1$ might be expected at some intermediate
$J_F$. However, it has been shown that in a N\'eel ordered bulk, the
correlations between the spins connected by the ferromagnetic bond remain
antiferromagnetic,\cite{sch94a} which is possible because of the broken
degeneracy of the triplet when it is coupled to an asymmetric environment.
Hence, counterintuitively, an $S_i=1$ behavior when $T\to 0$ may actually
never be realized with a ferromagnetic bond impurity. We will here consider
only the coupling strength $J_F=J$, for which the sign problem due to
frustration can be alleviated by a position averaging procedure, as
discussed in Sec.~III. Our QMC data shows that the impurity moment $S_i=0$
in this case, and the behavior is similar to the removed-bond impurity
model ($J_F=0$) shown in Fig.~\ref{fig2}(f).

Two-impurity models are useful for further clarifying the properties of the
single-impurity models. Here we will consider only the case of two vacancies,
which are chosen to be either at a fixed short distance from each other or
maximally separated on the $L\times L$ square lattices.

\subsection{Effective models}

The purpose of introducing effective models is to capture the dominant
mechanisms at play in the full Heisenberg models with very simple systems
containing a minimal number of adjustable parameters. The effective models are
constrained by two criteria: (i) they should reproduce the high-$T$ impurity
susceptibility, the sign of which depends on whether a spin has been added or
removed, and (ii) they should mimic the expected \cite{sachdev1} 
leading-order behavior $S_i^2/3T$ of the impurity susceptibilities at low $T$,
i.e., the alignment of the impurity moment with the local N\'eel order. 
Effective models are here considered for the full $S_i\neq 0$ impurity models 
shown in Fig.~\ref{fig1}.

In our effective models the local N\'eel order at the impurity is modeled by
a classical ``nonmagnetic'' vector $\mathbf{N}$. When a single spin
$\mathbf{S}_0$ is removed from the full model of the pure system, as shown to
the left in Fig.~\ref{fig1}(b), the remaining system has an $S=1/2$ ground
state due to the sublattice asymmetry. Hence, the effective model corresponding
to the vacancy system, shown to the right in Fig.~\ref{fig1}(b), is simply
defined with a single effective remnant ``environment'' spin-$\frac{1}{2}$
$\mathbf{S}_e$. This spin is coupled to a classical unit vector $\mathbf{N}$
representing the orientation of the local N\'eel order. The magnitude of
this order is absorbed in the coupling strength $r$. The effective model
for the pure system is naturally obtained by reinserting the spin
$\mathbf{S}_0$, as shown to the right in Fig.~\ref{fig1}(a), with $\alpha >0$
for antiferromagnetic coupling. The effective model for the added-spin system,
shown to the right in Fig.~\ref{fig1}(c), is obtained by coupling an extra
spin-$\frac{1}{2}$ $\mathbf{S}_a$ to $\mathbf{S}_0$. Finally, the effective
model for the system with a configuration of four ferromagnetic bonds, shown
to the right in Fig.~\ref{fig1}(d), is obtained by simply changing the sign
of $\alpha$, i.e., by making the coupling ferromagnetic instead of
antiferromagnetic. To summarize, the Hamiltonians of the effective models,
corresponding to the full models in Fig.~\ref{fig1}, are
\begin{subequations}
\begin{eqnarray}
H_{(a)}^{\text{eff}}&=&r\mathbf{N}\cdot \mathbf{S}_{e}+\alpha \mathbf{S}_{0}
        \cdot\mathbf{S}_{e},\label{eff_pure}\\
H_{(b)}^{\text{eff}}&=&r\mathbf{N}\cdot \mathbf{S}_{e},\label{eff_vac}\\
H_{(c)}^{\text{eff}}&=&r\mathbf{N}\cdot \mathbf{S}_{e}+\alpha \mathbf{S}_{0}
        \cdot\mathbf{S}_{e}+J_{\perp}\mathbf{S}_{a}\cdot \mathbf{S}_{0},
        \label{eff_add}\\
H_{(d)}^{\text{eff}}&=&r\mathbf{N}\cdot \mathbf{S}_{e}-\alpha \mathbf{S}_{0}
        \cdot\mathbf{S}_{e}.\label{eff_fer}
\end{eqnarray}
\label{effmodels}
\end{subequations}
The parameters $r>0$ and $\alpha >0$ cannot be derived in any trivial way.
The magnitude of $r$ should in principle depend on $T$, but the $T$ dependence
can be expected to be weak once the amplitude of the order has developed
locally close to the impurity. One could also argue that a direct coupling
between ${\bf N}$ and the central spin $\mathbf{S}_0$ should be included.
Such a coupling is clearly mediated through the four nearest neighbors of
$\mathbf{S}_0$. However, in the spirit of keeping the models as simple as
possible, we here chose to accomplish this coupling indirectly through the
remnant environment spin $\mathbf{S}_e$. One can further anticipate that the
optimum values for the couplings $r$ and $\alpha$ will depend on the impurity
type, since the effective impurity spin is spread out and its coupling
is mediated through the local environment of $\mathbf{S}_0$, which will be
distorted in different ways by different impurities. However, we will show
that the same parameters, $r/J\approx 1.90$ and $\alpha /J\approx 2.25$,
\cite{sepnote2} actually give an overall reasonable agreement for all the
$S_i>0$ impurity types considered here.

The procedure for determining the susceptibilities of the effective models
is straightforward. An external applied field $\mathbf{h}=h_z\mathbf{e}_z$
defines the $z$ direction. The magnetization operators $\mathbf{M}_{(i)}$,
corresponding to the effective Hamiltonians in Eqs.~(\ref{effmodels}),
have the $z$ components $M_{(a)}^{z}=S_{0}^{z}+S_{e}^{z}$, $M_{(b)}^{z}=
S_{e}^{z}$, $M_{(c)}^{z}=S_{a}^{z}+ S_{0}^{z}+S_{e}^{z}$, and $M_{(d)}^{z}=
M_{(a)}^{z}$. The susceptibilities are given by the usual formula
\begin{eqnarray}
\chi ^{z}_{(i)}&=&\frac{\partial \bm{\langle}\langle M_{(i)}^{z}
\rangle \bm{\rangle}_{\mathbf{N}}}{\partial h_{z}}|_{h_{z}=0}
\label{kubo}\\
&=&\int _{0}^{1/T}d\tau \bm{\langle}\langle M_{(i)}^{z}(\tau)M_{(i)}^{z}(0)
\rangle \bm{\rangle}_{\mathbf{N}}-\frac{1}{T}\bm{\langle}\langle M_{(i)}^{z}
\rangle \bm{\rangle}_{\mathbf{N}}^2,\nonumber
\end{eqnarray}
where $i=a,b,c$, and $d$. Here the inner brackets $\bm{\langle}\cdot
\bm{\rangle}$ indicate the quantum mechanical expectation value for a fixed
direction of $\mathbf{N}$, and $\bm{\langle}\cdot \bm{\rangle}_{\mathbf{N}}$
denotes the classical orientation average. The imaginary-time evolved operator
$M_{(i)}^{z}(\tau)=\exp (\tau H_{(i)}^{\text{eff}})M_{(i)}^{z}\exp
(-\tau H_{(i)}^{\text{eff}})$. Expressing $M_{(i)}^{z}$ in the coordinate
system defined by $\mathbf{N}$,
\begin{equation}
M^z_{(i)} = \cos{(\Theta)} M^\parallel_{(i)} - \sin{(\Theta)}M^\perp_{(i)},
\end{equation}
where $\parallel$ and $\perp$ denote the directions parallel and perpendicular
to $\mathbf{N}$, the expectation values can be easily calculated. The second
term in Eq.~(\ref{kubo}) vanishes. The first term can be separated to
components:
\begin{equation}
\chi ^{z}_{(i)}=\frac{1}{3}\chi ^{\parallel}_{(i)}+\frac{2}{3}\chi
^{\perp}_{(i)},
\label{usdef}
\end{equation}
where the prefactors originate from the classical orientation averaging. The
susceptibility components are
\begin{subequations}
\begin{eqnarray}
\chi^{\parallel}_{(i)}&=&\int_{0}^{1/T}d\tau \langle
        M_{(i)}^{\parallel}(\tau)M_{(i)}^{\parallel}(0)\rangle =
        \frac{1}{T}\langle (M_{(i)}^{\parallel})^2 \rangle ,~~
        \label{linres1} \\
\chi^{\perp}_{(i)}&=&\int_{0}^{1/T}d\tau \langle
        M_{(i)}^{\perp}(\tau)M_{(i)}^{\perp}(0)\rangle ,~~
        \label{linres2}
\end{eqnarray}
\end{subequations}
which can be easily evaluated in the $\parallel$ basis. In this basis
Eq.~(\ref{linres1}) has the simple form because $[H_{(i)}^{\text{eff}},
M_{(i)}^{\parallel}]=0$.

\begin{figure}
\includegraphics[width=8cm,clip]{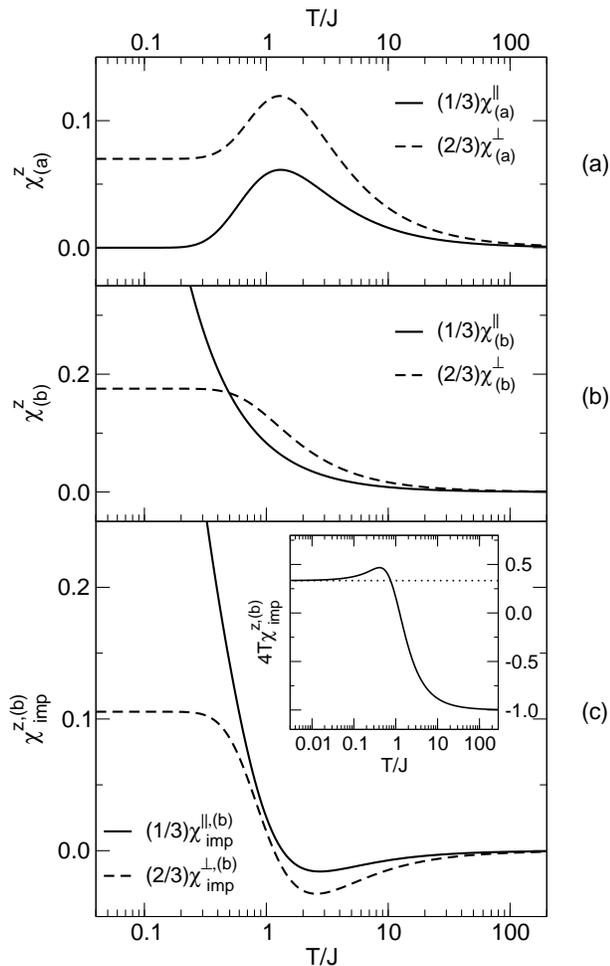}
\caption{The components of the susceptibilities of the effective models
for the pure (a) and vacancy (b) systems. The components of the impurity
susceptibility $\chi ^{z,(b)}_{\text{imp}}=\chi ^{z}_{(b)}-\chi ^{z}_{(a)}$
are shown in (c). The inset shows the $4T\chi _{\text{imp}}^{z,(b)}\sim 1/3$
behavior as $T\to 0$.}
\label{fig3}
\end{figure}

A dominant feature of the effective models is the alignment of a 
quantum spin with a classical vector. This can be appreciated by examining the 
simple effective Hamiltonian $H^{\text{eff}}_{(b)}$, given in 
Eq.~(\ref{eff_vac}), for the vacancy system. The corresponding 
susceptibility is given by
\begin{eqnarray}
\chi _{(b)}^{z}&=&\frac{1}{3}\chi _{(b)}^{\parallel}+\frac{2}{3}
                 \chi _{(b)}^{\perp}\nonumber \\
&=&\frac{1}{3}\frac{1}{4T}+\frac{2}{3}\frac{1}{2r}\tanh \left
                 (\frac{r}{2T}\right ).
\label{analytic}
\end{eqnarray}
The temperature dependence of the two components is graphed in
Fig.~\ref{fig3}(b) for $r=1.90$. Since $S_e=1/2$ in this model,
the $\parallel$ component can be written as
$(1/3)\chi ^{\parallel}_{(b)}={S_{e}}^2/3T$. Hence, the classical Curie
prefactor ${S_e}^2$, instead of the usual quantum mechanical prefactor
$S_e(S_e+1)$, is a consequence of the finite coupling between the spin
$\mathbf{S}_e$ and the classical vector $\mathbf{N}$. This is precisely
the low-$T$ leading order behavior proposed\cite{sachdev1} in
Eq.~(\ref{subir}) in the renormalized classical regime of a 2D
antiferromagnet. The perpendicular component in Eq.~(\ref{analytic}) tends
to a constant at low $T$. On the other hand, in the limit $r\to 0$, i.e.,
when $\mathbf{S}_e$ and $\mathbf{N}$ decouple, $\mathbf{S}_e$ recovers its
quantum identity and the susceptibility has the usual Curie form
$\chi ^{z}_{(b)}\to S_e(S_e+1)/3T$.

The effective models also serve the purpose of elucidating the steps in
determining the impurity susceptibility $\chi ^{z,(b)}_{\text{imp}}=
\chi ^{z}_{(b)}-\chi^{z}_{(a)}$. The separation in components of the 
susceptibility for the effective pure system, $\chi^{z}_{(a)}$, is shown in 
Fig.~\ref{fig3}(a). The parallel component $(1/3)\chi ^{\parallel}_{(a)}$ 
vanishes at low $T$ since the spins $\mathbf{S}_0$ and $\mathbf{S}_e$ are then
aligned antiferromagnetically with respect to $\textbf{N}$. The perpendicular 
component $(2/3)\chi ^{\perp}_{(a)}$ assumes a constant value at low $T$. The 
components of the susceptibility $\chi^{z}_{(b)}$ for the vacancy system were 
given analytically in Eq.~(\ref{analytic}) and are shown in 
Fig.~\ref{fig3}(b). Finally, the components of the impurity
susceptibility are shown in Fig.~\ref{fig3}(c). At high $T$, the impurity
susceptibility is just the sum of the Curie contributions of each independent
spin, i.e., $\chi _{\text{imp}}^{z,(b)}\to 1/4T-2/4T=-1/4T$. At low $T$ the
parallel component diverges, while the perpendicular component becomes a
constant. The inset verifies that the parallel component is responsible
for the ${S_e}^2/3T$ behavior, since $4T\chi _{\text{imp}}^{z,(b)}\sim 1/3$
as $T\to 0$. Besides being capable of reproducing the expected low-$T$
leading order behavior of the full models, the effective models also account
quite accurately for impurity specific behavior at intermediate $T$, as will
be shown in Sec.~\ref{sec_results}. There we also demonstrate that the
effective models account accurately for the $T$ dependence of the
internal energy.

\section{Quantum Monte Carlo method}\label{sec_method}

The numerical method employed here for the full Heisenberg models is the
operator-loop formulation of the stochastic series expansion (SSE) QMC method.
It is a method based on importance sampling of the terms of the Taylor
series for the density matrix. Its application to the Heisenberg model has
been discussed in detail, e.g., in Refs.~\onlinecite{sse1} and
\onlinecite{sse2}. The method is only briefly outlined here in order to
discuss some important aspects of the impurity work, including a trick
for alleviating the sign problem for the frustrated-bond system and the use
of improved estimators for reducing statistical errors.

The Hamiltonian of the Heisenberg antiferromagnet in Eq.~(\ref{pure})
can be cast into the form
\begin{equation}
H_{(a)}=-\frac{J}{2}\sum_{b=1}^{N_b}(H_{1,b}-H_{2,b})+\frac{JN_b}{4},
\label{ham1}
\end{equation}
where the operators $H_{1,b}$ and $H_{2,b}$, defined by
\begin{subequations}
\begin{eqnarray}
H_{1,b}=2\left (\frac{1}{4}-S^z_{i(b)}S^z_{j(b)}\right ),\label{h1b} \\
H_{2,b}=S^+_{i(b)}S^-_{j(b)}+S^-_{i(b)}S^+_{j(b)}, \label{h2b}
\end{eqnarray}
\label{ham2}
\end{subequations}
are diagonal and off-diagonal, respectively, in the basis
$\{|\alpha \rangle =|S_1^z,S_2^z,\ldots,S_N^z\rangle \}$ used in the
simulation.
An exact expression for the partition function $Z$ is obtained by expanding 
the density matrix $e^{-\beta H}$ in a Taylor series at inverse temperature 
$\beta =1/T$ ($k_B=1$). The series can be truncated at some expansion power 
$n_{\rm max}=M$, since terms of order greater than $n \propto N\beta$ give 
an exponentially vanishing contribution.\cite{sse1} The truncated partition 
function is then given by 
\begin{equation}
Z=\sum _{\alpha}\sum _{S_M}W(\alpha ,S_M)\left \langle \alpha \left
|\prod _{i=1}^M H_{a_i,b_i}\right |\alpha \right \rangle .
\label{part}
\end{equation}
Since the matrix element of the operator product takes the values
0 and 1, the statistical weight of a contributing configuration is\cite{sse1}
\begin{equation}
W(\alpha ,S_M)=\frac{(-1)^{n_2}(\beta J)^n(M-n)!}{2^nM!}.
\label{weight}
\end{equation}
A number of $M-n$ identity operators $H_{0,0}=I$ have been inserted
in the matrix element of each term in Eq.~(\ref{part}), with expansion
order $n<M$, and the change in prefactor reflects the number of different
ways to distribute the $n$ Hamiltonian operators among the $M$ positions.
The symbol $S_M$ denotes a sequence of operator indices,
\begin{equation}
S_M=(a_1,b_1),(a_2,b_2),\ldots ,(a_M,b_M),
\label{seq}
\end{equation}
where $a_i\in \{1,2\}$ and $b_i\in \{1,\ldots ,N_b\}$,
corresponding to the operators $H_{a_i,b_i}$ in Eqs.~(\ref{ham2}),
or $(a_i,b_i)=(0,0)$, corresponding to the identity operator
$H_{0,0}$. For a given sequence $S_M$ the order $n$ then
denotes the number of non-$(0,0)$ operators in the sequence.
For a nonfrustrated lattice, the number $n_2$ of off-diagonal
operators $(2,b_i)$ in the sequence $S_M$ is always even for 
nonvanishing contributions, thus yielding a positive definite
statistical weight $W(\alpha ,S_M)$ in Eq.~(\ref{weight}).

With a positive definite expansion, the partition function $Z$ can be
stochastically evaluated by importance-sampling in the configuration
space $(\alpha ,S_M)$. For this purpose an algorithm consisting of
two different configuration updates is used. In the first update
(diagonal update) the sequence $S_M$ is traversed from beginning to end,
while attempting substitutions $(0,0)\leftrightarrow(1,b_i)$. The
substitution $(0,0)\rightarrow(1,b_i)$ is attempted only if the spins
connected by bond $b_i$ are antiparallel [for a nonvanishing contribution
with the definition of the diagonal operator in Eq.~(\ref{h1b})]. The
probabilities to use for accepting/rejecting the change have been given
elsewhere, e.g., in Ref.~\onlinecite{sse2}. An accepted attempt changes the
expansion order $n$ by $\pm 1$. If an off-diagonal operator $(2,b_i)$ is
encountered no single operator substitution can be carried out, and
instead the saved state $|\alpha \rangle$ is updated by flipping the two
spins connected by the bond $b_i$, so that the state on which the diagonal
operators act are always available when attempting and update
$(0,0)\rightarrow(1,b_i)$. In the second update (operator-loop update) the
sequence $S_M$ is uniquely decomposed into a number $N_l$ of operator loops,
in which substitutions $(1,b_i)\leftrightarrow(2,b_i)$ can be carried out, 
independently with probability $1/2$ for each loop. All the spins associated 
with the loops are also flipped. During the operator-loop update the order 
$n$ is kept fixed and the weight of the configuration is unchanged. The
operator-loop update was introduced and discussed in detail in
Ref.~\onlinecite{sse2}.

The simulation is started with a random state $|\alpha \rangle$ and
an empty sequence $S_M=(0,0),(0,0),\ldots ,(0,0)$ of arbitrary (short) length
$M$. One Monte Carlo step (MC step) consists of a diagonal update followed
by an operator-loop update. During the equilibration stage of the simulation 
the cutoff $M$ is adjusted to always exceed the maximum order $n$ reached. 
Hence, the truncated partition function $Z$ in Eq.~(\ref{part}) is no 
approximation. Observables are measured after every MC step and expectation 
values and their errors are determined by the usual method of data binning.
Estimators for various observables of the Heisenberg antiferromagnet, in the 
context of the SSE method, are discussed in Ref.~\onlinecite{sse1}. The 
susceptibility is given in Eq.~(\ref{chidef}), where the sum is evaluated 
in the stored state $|\alpha\rangle$. The internal energy and the specific 
heat are given by\cite{sse1}
\begin{subequations}
\begin{eqnarray}
E&=&-\frac{\langle n\rangle}{\beta},\label{erg}\\
C&=&\langle n^2\rangle-\langle n\rangle^2-\langle n\rangle.\label{spc}
\end{eqnarray}
\label{est}
\end{subequations}

The operator-loop formulation of the SSE method, as described above,
is directly applicable to the isotropic Heisenberg antiferromagnet.
Impurities in the form of vacancies, added spins, and missing bonds can
be included with only very minor changes in the algorithm. In the added-spin 
impurity case, the only change in a program for the pure model is that
the acceptance probabilities in a diagonal update $(0,0)\rightarrow(1,b_k)$, 
involving the additional bond $k$ connecting the impurity spin, depend on 
the bond strength $J_k$. However, the impurities consisting of a frustrating 
ferromagnetic bond or four nonfrustrating ferromagnetic bonds necessitate 
some additional considerations, as will be discussed next.

For a ferromagnetic bond, the diagonal bond operator (\ref{h1b}) is defined 
as $2(1/4+S^z_iS^z_j)$ and the off-diagonal (\ref{h2b}) is multiplied by $-1$.
During the diagonal update the substitution $(0,0)\rightarrow(1,b_i)$, where 
$b_i$ is a ferromagnetic (antiferromagnetic) bond, is hence attempted only if 
the spins connected by bond $b_i$ are parallel (antiparallel). The rules for
constructing the operator loops are also modified, as discussed in
Ref.~\onlinecite{sse3}. For the impurity consisting of four nonfrustrating
ferromagnetic bonds, the expression for the statistical weight $W$ in 
Eq.~(\ref{weight}) is still valid if $n_2$ is replaced by the number 
$n_{A2}$ of off-diagonal operators $(2,b_i)$ acting on antiferromagnetic 
bonds. Because of the symmetry of the arrangement of four ferromagnetic 
bonds, this number also has to be even, and, therefore, the weight $W$ is still
positive definite. 

A single frustrating ferromagnetic bond (here with coupling $J_F=J$) in 
the Heisenberg antiferromagnet gives rise to a sign problem. Proceeding as
in the case of four ferromagnetic bonds discussed above, the sign would be 
determined by the number of spin flips of antiferromagnetic bonds, which 
now can be even or odd. Since the total number of flips still has to be even,
we can also define the sign as $(-1)^{n_{F2}}$, where $n_{F2}$ is the number 
of spin flips on the ferromagnetic bond. However, we can also proceed in a 
different way which allows for an alleviation of the sign problem by 
position-averaging when $J_F = J$. We then treat the ferromagnetic bond 
in the same way as an antiferromagnetic bond in the diagonal update, i.e., 
a diagonal operator can appear only on antiparallel spins. The sign will 
then be given by $(-1)^{n_F}$, where $n_F$ is the total number of
operators---diagonal and off-diagonal---operating on the ferromagnetic
bond. The simulation of the system with a ferromagnetic bond then 
proceeds exactly as the simulation of the pure antiferromagnet, i.e.,
expectation values can be calculated using $|W|$ and by reweighting the 
measurements with the sign $S=(-1)^{n_F}$ of the corresponding configuration,
\begin{equation}
\langle A\rangle =\frac{\langle AS\rangle _{|W|}}{\langle S\rangle
_{|W|}}.
\label{reweight1}
\end{equation}
In practice, however, the calculations become impossible when
$\langle S\rangle _{|W|}$ approaches zero. Here a technique based on 
positional averaging is used to tackle this problem. The idea is to replace 
the sign $S$ of a given configuration with the averaged 
sign\cite{sse4}
\begin{equation}
\Sigma =\frac{1}{N_b}\sum _{R}S(R),
\label{ave_sgn}
\end{equation}
where an average of the sign $S(R)=(-1)^{n_{F}(R)}$ is taken with respect
to all possible locations $R$ of the ferromagnetic bond. Expectation values 
are then given by
\begin{equation}
\langle A\rangle =\frac{\langle A\Sigma \rangle _{|W|}}{\langle \Sigma
\rangle_{|W|}}.
\label{rewieght2}
\end{equation}
This technique was discussed in a more general context in  
Ref.~\onlinecite{sse4}, where it was shown that it significantly alleviates 
the sign problem of the antiferromagnet with randomly positioned ferromagnetic
bonds. This came at the price of an approximation corresponding to switching 
to an ``annealed'' disorder. Here, in the single-impurity problem, there is
no approximation as the trick simply corresponds to simultaneously studying
systems with all possible locations of the ferromagnetic bond. When considering
only a single position $R$ of the ferromagnetic bond, the sign problem will
be more severe than with a redefinition of the diagonal operator discussed
above for the four-bond impurity. However, when using the position averaging
there will be some system size above which the statistic is improved. We here
obtained expectation values with reasonable statistical errors for system 
sizes up to $L=32$ at temperatures down to $T=J/8$. Since the evaluation of
the sign during the simulation is completely separate from the sampling
procedures, the effect of a ferromagnetic bond can actually be obtained as
a ``bonus'' while simulating the pure antiferromagnet. A drawback of the
position averaging method is that it does not allow for ferromagnetic bond
strengths $J_F\not =J$, except perhaps for $J_F$ very close to $J$ where
reweighting should work.

\begin{figure}
\includegraphics[width=8cm,clip]{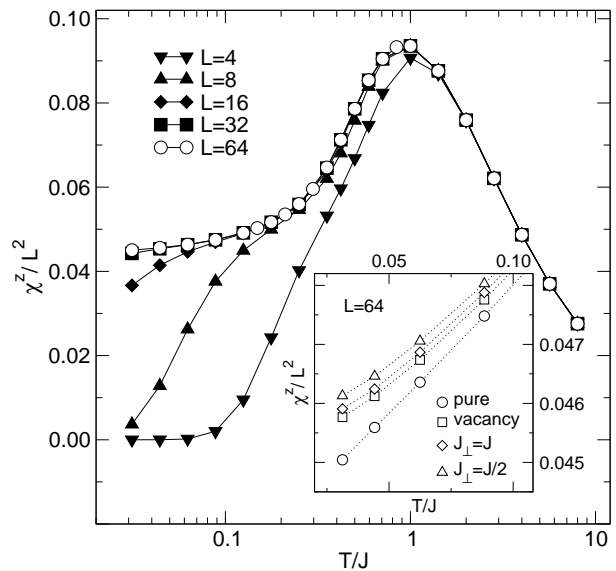}
\caption{SSE results for the magnetic susceptibility 
of the pure 2D Heisenberg
antiferromagnet for different system sizes $L$. Error bars are smaller than
the symbols. The inset shows a comparison between the low-$T$ behavior
for the pure, vacancy, and added-spin models with $L=64$. For the
added-spin model, two values on the coupling constant $J_\perp/J$
are considered.}
\label{fig4}
\end{figure}

We next briefly comment on the accuracy needed to study the impurity effects 
and the use of improved estimators for increasing the accuracy. 
For large $L$, the effect
of a single impurity on the magnetic susceptibility $\chi ^z$ is very small, 
as shown in the inset of Fig.~\ref{fig4}. In order to get acceptable errors 
for the impurity susceptibilities $\chi ^z_{\text{imp}}$ in Eq.~(\ref{chi_imp})
very precise values for the individual susceptibilities are clearly necessary.
To achieve this, an improved estimator \cite{evertz} is used. The general idea
is to reduce the statistical errors by replacing the value of an observable 
$A$ corresponding to a given Monte Carlo configuration by an estimator
$\bar A_i$ obtained by averaging over many equal-weight configurations during
the operator-loop update. In the case of the susceptibility, this is 
particularly simple since the magnetization is a conserved quantity.
Some of the loops will go through (once or multiple times) the state 
$|\alpha \rangle$, i.e., the state on which the ordered operator product is 
acting on in Eq.~(\ref{part}). Defining $\sigma_{i}^{z}$ as the sum over all
the spins in $|\alpha \rangle$ covered by the $i$:th loop, we clearly have
\begin{equation}
M^z=\sum_{i=1}^NS_{i}^{z}\equiv \sum_{i=1}^{N_l}\sigma_{i}^{z}.
\label{mag}
\end{equation}
We can now average this over all the $2^{N_l}$ ways of flipping the loops,
giving
\begin{equation}
\chi ^z=\beta \left \langle \sum_{i=1}^{N_l}{(\sigma_{i}^{z})}^{2}
\right \rangle .
\label{impr_usc}
\end{equation}
Figure~\ref{fig4} shows size-normalized results for the magnetic 
susceptibility obtained this way for the pure 2D Heisenberg antiferromagnet,
as well as low-$T$ data for systems with an impurity. We believe that these
results are the most accurate ones currently available for this model
and therefore also list selected numerical data in the Appendix. For the
internal energy and the specific heat, Eqs.~(\ref{erg}) and (\ref{spc}), no
improved estimator of the type discussed above can be constructed. The energy
can nevertheless be calculated to high accuracy, as also shown in the
Appendix. For the specific heat, it is very difficult to reach
good accuracy at low temperature. Nevertheless, we are able to clearly
discern the expected\cite{hasenfratz} behavior $C \propto T^2$ at low $T$,
as shown in Fig.~\ref{fig14} in the Appendix.

\section{results}\label{sec_results}

Here, in Sec.~\ref{subsec_spin}, we begin by presenting susceptibility
results for the $S_{i}\neq 0$ impurities illustrated in Fig.~\ref{fig1}. In
Sec.~\ref{subsec_3D} we consider the case of a vacancy in a 3D system,
and in Sec.~\ref{subsec_two} we look at the system with two vacancies.
We discuss results for the $S_{i}=0$ bond impurities (Fig.~\ref{fig2}) in
Sec.~\ref{subsec_bond}. In Sec.~\ref{subsec_other} we summarize our results
for the impurity effects on the energy and the specific heat.

\subsection{$S_{i}\neq 0$ impurities}\label{subsec_spin}

\begin{figure}
\includegraphics[width=8cm,clip]{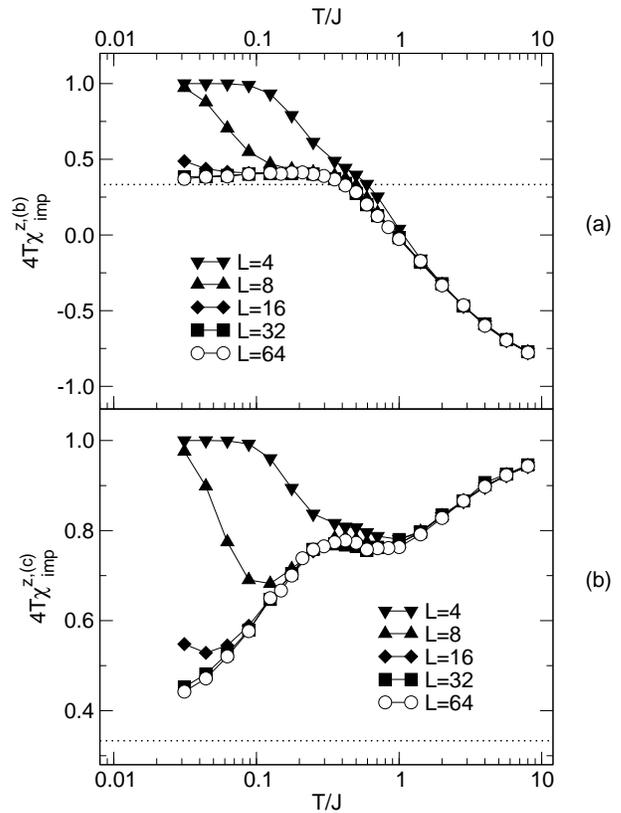}
\caption{The impurity susceptibilities of the (a) vacancy and (b)
added-spin ($J_{\perp}=J$) models, for different system sizes $L$. The dotted
lines show the expected asymptotic behavior $4T\chi^z_{\rm imp} \to 1/3$ as
$T\to 0$. Error bars are smaller than the symbols.}
\label{fig5}
\end{figure}

The impurity susceptibilities for a vacancy and an added spin with
$J_\perp =J$ are shown in Figs.~\ref{fig5}(a) and \ref{fig5}(b),
respectively. The results are multiplied by $4T$. At high $T$ the data
for different system sizes $L$ coincide, while at lower $T$ finite-size
effects are clearly seen for $L\leq 16$. The finite-size effects are due
to the $S=1/2$ ground states of the vacancy and added-spin models, to which
the system converges below an $L$ dependent crossover temperature, as has
recently been discussed by Sushkov.\cite{sushkov} For the largest system
size considered here, $L=64$, all finite-size effects are eliminated within
statistical errors for temperatures down to $T/J=1/32$. The observed behavior
at high $T$ for both impurity types is due to the fact that the total
susceptibility is then just the sum of the Curie contributions of each
independent spin, i.e.,
\begin{eqnarray}
\chi _{\text{imp}}^{z,(b,c)}&=&\chi _{(b,c)}^{z}-\chi _{(a)}^{z}
\nonumber \\
&\to&\frac{L^2\mp 1}{4T}-\frac{L^2}{4T}=\mp \frac{1}{4T}
\label{high}
\end{eqnarray}
as $T\to \infty$. The minus (plus) sign is for the vacancy (added-spin)
impurity model. According to the expression in Eq.~(\ref{subir}), the leading 
order behavior of the impurity susceptibility is $4T\chi _{\text{imp}}^{z}
\sim 4S_{i}^2/3$ as $T\to 0$. For a $S_{i}=1/2$ impurity, the constant value
$1/3$ should then be approached at low $T$. This is also clearly observed in
the size-converged ($L=64$) data for the vacancy impurity, shown in
Fig.~\ref{fig5}(a). For the added-spin impurity shown in Fig.~\ref{fig5}(b),
an approach of $4T\chi _{\text{imp}}^{z,(c)}$ to $1/3$ is also likely,
although the convergence occurs at lower $T$ than for the vacancy. At 
intermediate $T$ the results for the two different impurity types are 
strikingly different. Specifically, the shoulder-like structure with a 
minimum around $T/J\approx 0.8$ observed in the added-spin data has no 
counterpart in the vacancy data, but in both cases there is a maximum 
at $T/J\approx 0.2$. Some of the differences clearly are related to the
different $T \to \infty$ behaviors.

\begin{figure}
\includegraphics[width=8cm,clip]{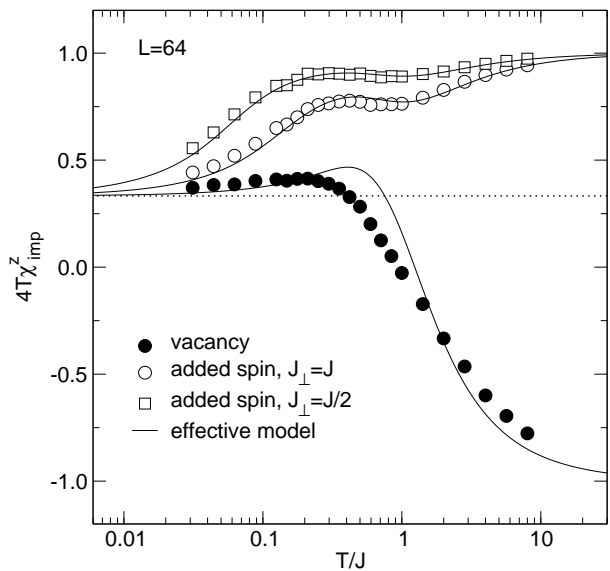}
\caption{$L=64$ results for the impurity susceptibilities of the
vacancy and added-spin models is compared to the results of the
corresponding effective models. The dotted line shows the value $1/3$.}
\label{fig6}
\end{figure}

In Fig.~\ref{fig6} the size-converged SSE data are compared with the
results of the effective models. Results are also shown for the
added-spin impurity with $J_{\perp}=J/2$. The values of the two parameters 
of the effective models, $\alpha /J=2.25$ and $r/J=1.90$,\cite{sepnote2} 
were chosen for optimal overall agreement between the SSE
data and the effective model results, for both the vacancy and the
added-spin systems. For this choice of values, the effective models reproduce
the added-spin data with a remarkable precision down to $T/J\approx 0.1$,
for both $J_{\perp}=J$ and $J_{\perp}=J/2$. Moreover, with the same set
of values a reasonable agreement is also obtained for the
vacancy system. Hence, the same parameters describe well a wide range of
coupling strengths to the added spin (the vacancy corresponds to
$J_\perp /J=\infty$).

In each of the three cases shown in Fig.~\ref{fig6}, the effective
models also reproduce the low-$T$ leading-order behavior suggested in
Eq.~(\ref{subir}), i.e., $4T\chi _{\text{imp}}^{z}\sim 4S_{i}^2/3=1/3$ for
a $S_{i}=1/2$ impurity. Hence, the effective models clearly contain the
dominant impurity physics and are able to distinguish between different 
impurity types in a broad $T$ range. In analogy with the results for the
effective models, the observed low-$T$ leading-order behavior of the full
Heisenberg models is ascribed to a susceptibility component parallel to a
locally N\'eel ordered domain coupled to the impurity, i.e.,
$(1/3)\chi _{\text{imp}}^{\parallel ,(i)}\sim S_{i}^2/3T=1/12T$, where $i=b,c$,
and $S_{i}=1/2$.

\begin{figure}
\includegraphics[width=8cm,clip]{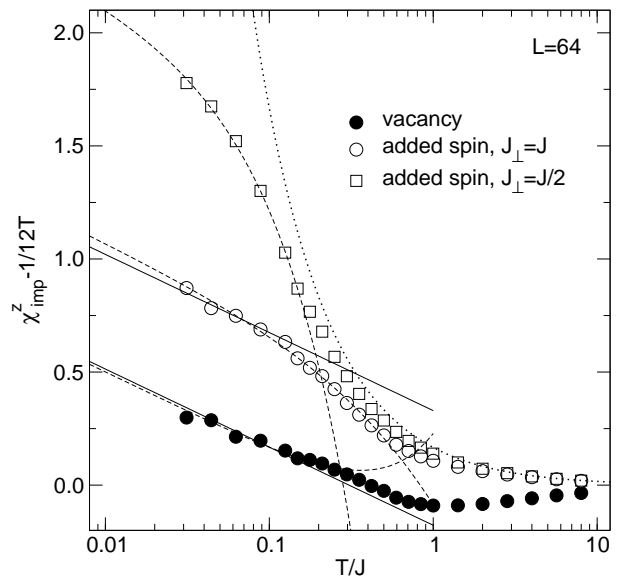}
\caption{SSE data for $\chi _{\text{imp}}^{z}-1/12T$ of the vacancy and
added-spin models with system sizes $L=64$. Straight lines and dashed
curves are fits of the theoretical results in Ref.~\onlinecite{sachdev2}
to our low-$T$ simulation data. The dotted curve shows the $1/6T$ behavior.}
\label{fig7}
\end{figure}

We next examine the thermodynamic low-$T$ impurity susceptibilities more
closely by subtracting from them the leading-order term $S_{i}^2/3T$. The
resulting quantities should then describe the transverse impurity
susceptibilities at low $T$, i.e., $(2/3)\chi _{\text{imp}}^{\perp ,(i)}
\sim \chi _{\text{imp}}^{z,(i)}-S_{i}^2/3T$.\cite{sepnote1} The results in
Fig.~\ref{fig7} for $\chi _{\text{imp}}^{z,(b,c)}-1/12T$ of the vacancy
impurity, and of the added-spin impurity with $J_{\perp}=J$, show an
apparent logarithmically divergent behavior as $T\to 0$. The results for the 
added-spin impurity with $J_{\perp}=J/2$ are not conclusive in this regard, 
but a similar log divergent behavior at still lower temperatures is 
clearly plausible. As 
$J_{\perp}/J\to \infty$, the magnetic properties of the
added-spin model should become equivalent to those of the 
vacancy model. In the limit $J_{\perp}/J\to 0$, on the other hand, the added
spin is decoupled from its host and the impurity susceptibility becomes
simply the susceptibility of a single spin ($1/4T$), i.e, $\chi _{\text{imp}}
^{z}-1/12T\sim 1/6T$. When comparing the SSE results in Fig.~\ref{fig7}
with each other, it then seems that the log divergent behavior starts at 
higher $T$ as the magnitude of the coupling to the added spin, $J_{\perp}/J$,
is increased. This can be naturally understood as an impurity moment
strongly coupled to the environment can develop only at $T$ below
$J_{\perp}$.

According to the theoretical expression by Sachdev and Vojta,\cite{sachdev2} 
Eq.~(\ref{subir}), the slopes of the the low-$T$ curves should be equal on
the log-linear scale used in Fig.~\ref{fig7}. The slope is given by
$S_{i}^2/3\pi \rho_{s}$, where $S_{i}$ is the ``bare'' impurity spin and
$\rho _s$ is the spin stiffness of the bulk-ordered antiferromagnet, for
which we use the value $\rho_{s}/J=0.181$.\cite{stiffnessvalue} Our results
for the vacancy and the $J_\perp =J$ added spin are indeed consistent with this
prediction. The straight solid lines are fits of the leading logarithmic 
part, $\propto \ln (C_1\rho_s/T)$, of Eq.~(\ref{subir}) to the low-$T$ data, 
whereas the dashed curves show fits including also the subleading correction
$\propto T\ln (C_2\rho_s/T)$. For the vacancy system we find $C_1\approx 1.7$
(in the leading-order fit) or $C_1\approx 1.6$ and $C_2\approx 0.3$, for 
the added-spin system ($J_{\perp}=J$) $C_1\approx 50$ or $C_1\approx 73$ 
and $C_2\approx 184$. For the added-spin impurity with $J_{\perp}=J/2$, no 
fit can be made with only the leading term, and we find 
$C_1\approx 10^5$ and $C_2\approx 10^{19}$.

\begin{figure}
\includegraphics[width=8cm,clip]{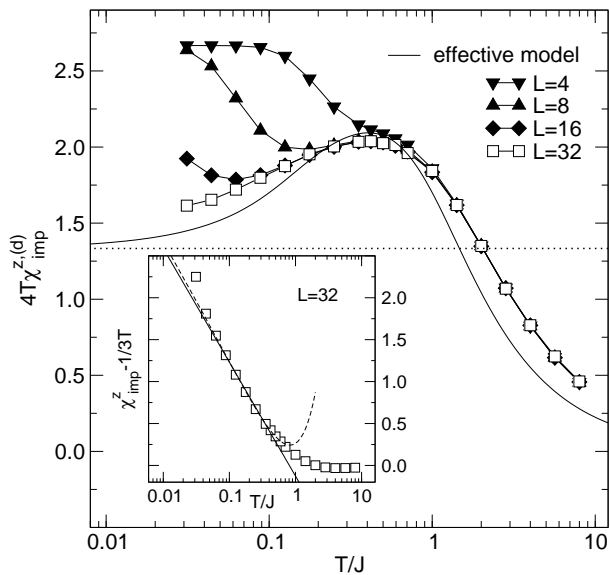}
\caption{Impurity susceptibilities for different system sizes $L$ of the
impurity model with four ferromagnetic bonds. The solid curve shows the
result of the corresponding effective model, which assumes the asymptotic
value $4/3$ (shown by the dotted line) at low $T$. The log divergent
behavior of $\chi _{\text{imp}}^{z}-1/3T$, for a system of size $L=32$, is
shown in the inset, where the solid line and the dashed curve are fits of
the theoretical results in Ref.~\onlinecite{sachdev2} to our low-$T$
simulation data.}
\label{fig8}
\end{figure}

Results for the impurity model with a configuration of four
ferromagnetic bonds are shown in Fig.~\ref{fig8}.
Again, the results for different $L$ coincide at high $T$, while
finite-size effects are seen at lower $T$. The high-$T$ observed behavior,
$4T\chi _{\text{imp}}^{z,(d)}\to 0$ as $T\to \infty$, is due to the fact that
the susceptibilities of the doped and the pure models cancel, since there is
an equal number of independent spins in both models at high temperatures. The
ground state spin of this model is $S=1$, and hence also an impurity moment
$S_i=1$ can be anticipated. The low-$T$ finite-size susceptibility should
then be $4T\chi ^{z}\sim 4T[S(S+1)/3T]=8/3$ and in the thermodynamic limit
$4T\chi ^{z}\sim 4T(S^2/3T)=4/3$. This behavior is indeed seen in 
Fig.~\ref{fig8}; for $L=4$ and $8$ the low-$T$  behavior dictated by
the ground state spin can be observed, while for $L=32$ the low-$T$
susceptibility is size-converged at least to $T/J=1/16$ and is consistent
with a convergence to $4/3$. We also show results for the corresponding
effective model. Using the same values for $\alpha /J$ and $r/J$ as 
previously for the vacancy and added-spin effective models, changing only 
the sign of $\alpha$, the behavior agrees qualitatively with the SSE results.
The inset of Fig.~\ref{fig8} shows $\chi _{\text{imp}}^{z,(d)}-1/3T$, which 
at low $T$ should be dominated by the transverse component 
$(2/3)\chi _{\text{imp}}^{\perp ,(d)}$. Again, an apparent log
divergent trend is observed. The straight line and the dashed curve are 
fits of Eq.~(\ref{subir}). Data for the two lowest $T$ are not included in 
these fit because of the finite-size effects that most likely remain 
here. Nevertheless, the results support the universal low-$T$ prefactor
(slope) of the leading logarithmic correction.

\begin{figure}
\includegraphics[width=8cm,clip]{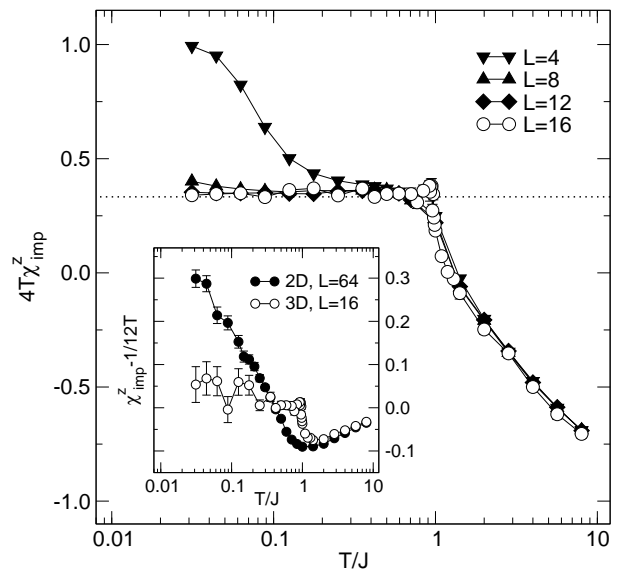}
\caption{The impurity susceptibilities of the 3D Heisenberg
antiferromagnet with a vacancy, for different system sizes $L$. The dotted
line shows the value $1/3$. A comparison between $\chi _{\text{imp}}^{z}
-1/12T$ of the 2D and 3D models is shown in the inset.}
\label{fig9}
\end{figure}

\subsection{Vacancy in a 3D system}\label{subsec_3D}

Here we discuss the case of a vacancy in the 3D Heisenberg antiferromagnet. 
Some predictions\cite{sachdev2} were recently made also for this system, 
but since we have not achieved sufficient accuracy they are not tested in 
detail here. The leading-order behavior can nevertheless be extracted. 
In Fig.~\ref{fig9} the SSE data are shown for different system sizes $L$
($N=L^3$), and a comparison between the 3D and 2D data is shown in the inset. 
The high-$T$ behavior, as well as the low-$T$ finite-size effects, have the 
same explanations as those given for the 2D results. For the largest system 
size, $L=16$, most finite-size effects are eliminated within
statistical errors in the $T$ range considered. The observed thermodynamic 
behavior is reminiscent of the 2D results in Fig.~\ref{fig5}(a), with the 
exception that the transition to a constant-valued behavior now occurs 
abruptly at $T/J\approx 0.95$, which is the N\'eel temperature $T_N$ of the 
model.\cite{sandvik3d} There are signs of a singular behavior of the
impurity susceptibility at the transition. At temperatures $T\leq T_N$, the 
susceptibility is seen to follow very closely the proposed
\cite{sachdev1,sachdev2} behavior $S^2/3T$. It should be noted that although 
3D order sets in below $T_N$, our finite-size systems nevertheless do not 
break the symmetry and the direction of the N\'eel vector is not fixed. In 
an infinite symmetry-broken system, the $S^2/3T$ behavior would not be 
present if the magnetization fluctuations are defined with respect to the
average in the direction of the fixed N\'eel vector.

In the inset of Fig.~\ref{fig9}, the perpendicular component
$(2/3)\chi _{\text{imp}}^{\perp}=\chi _{\text{imp}}^z-(1/3)\chi
_{\text{imp}}^{\parallel}$ is compared to the analogous quantity of the 2D
model. Although the statistical accuracy is not very high at low temperature,
it is clear that the behaviors are different. The 3D results do not indicate
any log divergent behavior of the type observed in the 2D system. Instead 
an almost constant behavior is observed, as also predicted in the field
theory.\cite{sachdev2}

\begin{figure}
\includegraphics[width=8cm,clip]{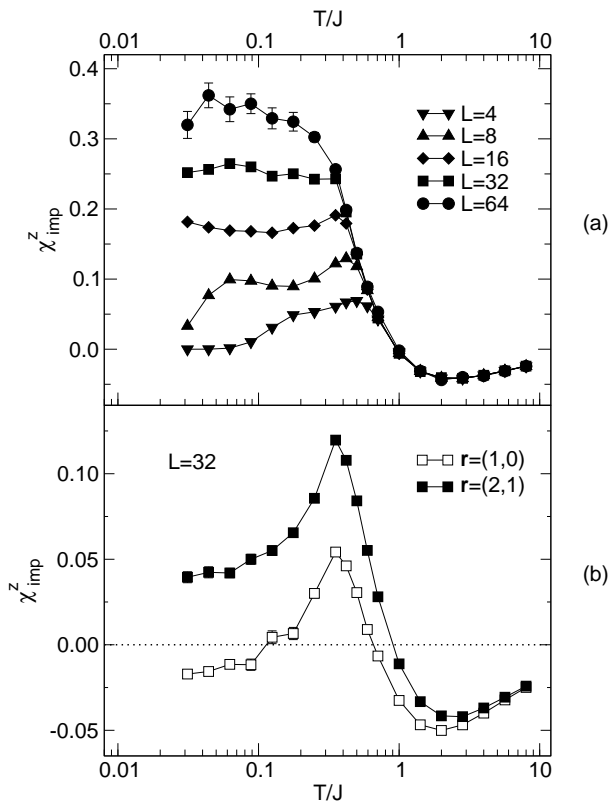}
\caption{Impurity susceptibility for different system sizes $L$ of
the square lattice with two vacancies. The vacancies are as far apart
as possible in (a). In (b) the $L=32$ results are shown for the cases
when the two vacancies are nearest neighbors (open symbols) and at
distance $\mathbf{r}=(2,1)$ from each other (solid symbols).}
\label{fig10}
\end{figure}

\subsection{Two vacancies in 2D}\label{subsec_two}

Next we present SSE results for the 2D Heisenberg antiferromagnet with two
vacancies on different sublattices. The results for the impurity
susceptibility are multiplied by a factor $1/2$, so that single-impurity
values should be obtained when the correlation length is much shorter
than the separation between the vacancies. When $T$ is lowered, interactions
between the impurities become important as the correlation length $\xi$
grows exponentially. At $T$ corresponding to a correlation length of the
same order as the vacancy separation, the moments due to the two vacancies
on different sublattices are pinned by the local N\'eel order antiparallel
to each other, resulting in a rapid quenching of the parallel component of
the impurity susceptibility. Hence, $\chi _{\text{imp}}^z$ does not diverge
as $T\to 0$. The data shown in Fig.~\ref{fig10}(a) is for the case of maximum
separation of two vacancies on different sublattices; $\mathbf{r}=(L/2-1,L/2)$.
Since $\xi$ diverges exponentially as $T\to 0$, the point at which
$\chi _{\text{imp}}^z$ deviates from the divergent single-vacancy behavior 
moves only very slowly to lower $T$ as $L$ is increased. For larger $L$, an
almost constant $\chi _{\text{imp}}^z$ is observed. However, no sign of
convergence of the plateau value is seen. Clearly, in a system of finite size
there will always be some interaction also between the perpendicular
components of the two vacancies, and hence even for large $L$ the two-vacancy
model does not trivially reproduce the single-vacancy results below some
temperature. It is plausible, however, that the roughly $\ln (L)$ divergence
of the plateau height seen in Fig.~\ref{fig10}(a) continues as $L\to \infty$.
This would be fully in line with the log divergent
$\chi _{\text{imp}}^{\perp ,(b)}$ for the single vacancy.

The very sudden crossover from divergent to almost $T$ independent
behavior seen in Fig.~\ref{fig10}(a) speaks for a component
$\chi _{\text{imp}}^{\perp ,(i)}$ aligning strongly to the local N\'eel
order (which becomes the global order at the $L$ dependent crossover
temperature), and justifies the separation into parallel and transverse
(with respect to the local fluctuating N\'eel vector) impurity
susceptibility components already at intermediate $T$. However, the
longitudinal component is not strictly $S_i^2/3T$; the recent field
theory by Sachdev and Vojta predicts that the remaining longitudinal
contributions, once this leading term has been subtracted, has a temperature
dependence $\propto T\ln{(1/T)}$. Nevertheless, the transverse contribution,
which is $\propto \ln{(1/T)}$ at low $T$, dominates.

In Fig.~\ref{fig10}(b), SSE data is shown for the case of the two vacancies
being nearest neighbors, $\mathbf{r}=(1,0)$, as well as at separation
$\mathbf{r}=(2,1)$ on the square lattice. Again, the single-vacancy data
are reproduced at high $T$. In contrast to the divergent trend seen in the
maximum-separation data in Fig.~\ref{fig10}(b), the finite-size behavior has
now converged to a near constant at low $T$, and no signs of a log divergence
as a function of $L$ is observed. In the figure we show only $L=32$ results,
which are almost converged to the thermodynamic limit. The absence of log
corrections for two vacancies at fixed separation is consistent with results
of a Green's function calculation,\cite{nagaosa} where the introduction of
a second extra spin destroyed the log divergence in the frequency dependent
$T=0$ susceptibility observed for the system with a single extra spin.

\subsection{$S_{i}=0$ impurities}\label{subsec_bond}

\begin{figure}
\includegraphics[width=8cm,clip]{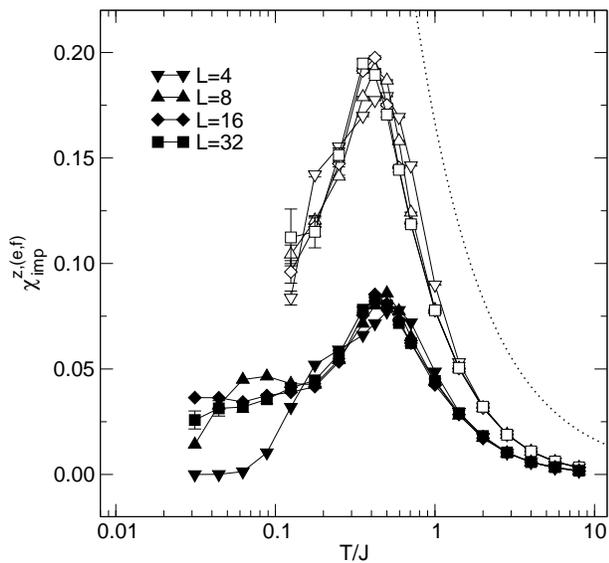}
\caption{Impurity susceptibilities for different system sizes $L$ of the
models with a missing bond (solid symbols) and a ferromagnetic bond with
$J_F=J$ (open symbols). The dotted curve shows the expected high-$T$
behavior $1/6T$ for large $J_F/J$.}
\label{fig11}
\end{figure}

We next turn to the QMC results shown in Fig.~\ref{fig11} for the 2D
Heisenberg antiferromagnet with a ferromagnetic bond or a missing bond,
i.e., with $J_F=J$ or $J_F=0$, respectively. In this case the impurity
susceptibilities do not diverge as $T\to 0$, and the results are not,
therefore, multiplied with $T$. The observed high-$T$ behavior of each
model, $\chi _{\text{imp}}^z\to 0$, is due to the fact that the
susceptibilities of the pure and the doped models cancel, since there is
an equal number of independent spins in both models at high temperatures.
For the missing-bond impurity, low-$T$ finite-size effects are clearly
seen for $L=4$ and $8$, while for the largest system size $L=32$, the
results should be almost size-converged and show little temperature 
dependence at low $T$. The observed finite-size behavior,
$\chi _{\text{imp}}^z(T\to 0)\to 0$ reflects the $S=0$ ground state, and
clearly the size-converged $T$ dependence also speaks for an $S_i=0$
impurity. Results for the ferromagnetic-bond impurity are limited to
temperatures down to $T/J=1/8$, because of the sign problem caused by the
frustrating ferromagnetic bond. The data are reminiscent of the missing-bond
results, and hence also the ferromagnetic-bond impurity has $S_i=0$.
Both models are, clearly, special cases of the system with one ferromagnetic
bond of arbitrary strength $J_F$. It would be interesting to investigate
how the impurity spin magnitude $S_i$ changes as $J_F$ is increased. For
$J_F/J\gg 1$, the two spins connected by the ferromagnetic bond form a
triplet and hence should give an $S_i=1$ Curie contribution
$S_i(S_i+1)/3T=2/3T$ when $J \alt T \alt J_F$. The remaining $N-2$ spins each
contribute $1/4T$, and hence the impurity susceptibility should be $1/6T$
in this regime. In Fig.~\ref{fig11} the results for $T > J$ are closer to
this form for $J_F=J$ than for $J_F=0$, but the requirement $J < T < J_F$
is not satisfied and the deviations (reduction relative to $1/6T$) reflect an
expected crossover from the high-$T$ independent-spin form
$\chi _{\text{imp}}^z \approx 0$.

An interesting question is whether the classical-like Curie behavior
$\chi _{\text{imp}}^z \approx S_i^2/3T$ with $S_i=1$ can be observed
in this model for $J_F > J$. As already discussed  in Sec.~II, the asymmetric
coupling to the bulk of the two spins connected by the ferro bond most likely
implies a $T \to 0$ behavior corresponding to $S_i=0$ for any finite $J_F$.
This is because an $S_i=1$ impurity requires that the two spins at the ferro
bond are dominantly in the $m^z=1$ state $|\uparrow\uparrow\rangle$ with
respect to the local N\'eel order (in a semiclassical picture such as
our effective impurity model), whereas in fact the couplings in this
case instead favor the $m^z=0$ component $(|\uparrow\downarrow\rangle + 
|\downarrow\uparrow\rangle) /\sqrt{2}$.\cite{sch94a}

\subsection{Internal energy and specific heat}\label{subsec_other}

\begin{figure}
\includegraphics[width=8cm,clip]{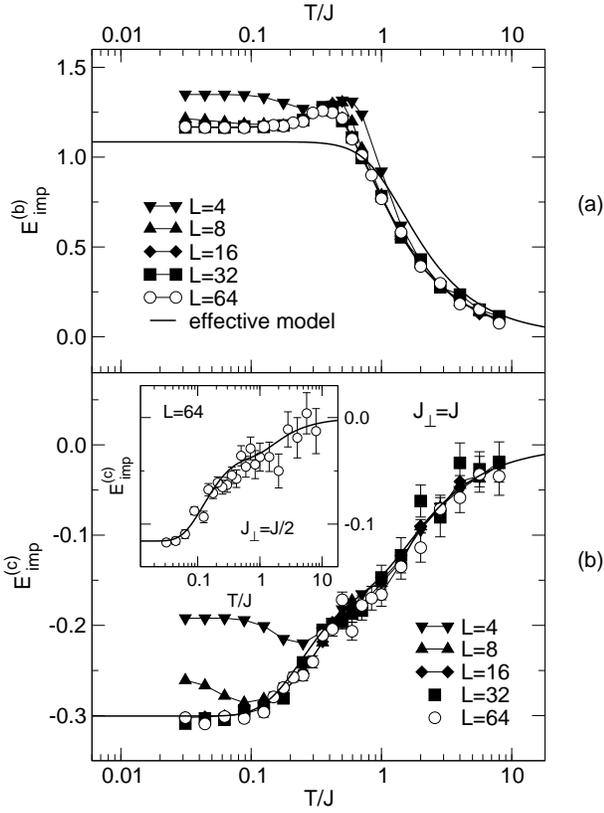}
\caption{The impurity energies of the (a) vacancy and (b)
added-spin ($J_{\perp}=J$) models, for different system sizes $L$.
The inset shows the $L=64$ QMC results for the added-spin model with
$J_{\perp}=J/2$. The curves are results of the corresponding effective
models.}
\label{fig12}
\end{figure}

We finally discuss our SSE calculations concerning impurity effects on the
internal energy and the specific heat, which we have obtained using
the estimators in Eqs.~(\ref{erg}) and (\ref{spc}), respectively. In
analogy to Eq.~(\ref{chi_imp}), we again define the impurity quantities
as differences between the doped and the pure systems, i.~e.,
\begin{subequations}
\begin{eqnarray}
E_{\text{imp}}^{(i)}=E_{(i)}-E_{(a)},\label{e_imp}\\
C_{\text{imp}}^{(i)}=C_{(i)}-C_{(a)},\label{c_imp}
\end{eqnarray}
\label{other_imp}
\end{subequations}
where $i=b,c,d,e$, and $f$, correspond to the different impurity
systems shown in Figs.~\ref{fig1} and \ref{fig2}, and the symbol $a$
denotes the pure system. In Fig.~\ref{fig12} results for the impurity
energies are shown for the vacancy model (a) and the added-spin model (b)
with $J_{\perp}=J$ and $J_{\perp}=J/2$ (shown in the inset). At high $T$
the impurity energies vanish, since the mean energy of each independent spin
becomes zero. For $L=64$, all finite-size effects are eliminated within
statistical errors for both models in the $T$ range considered. Since the
vacancy system has four antiferromagnetic bonds less than the pure system,
the impurity energy $E_{\text{imp}}^{(b)}$, shown in Fig.~\ref{fig12}(a),
is positive at all $T$. At low $T$ the results converge to a constant value,
which should be equal to the energy cost of removing one spin from an
infinite lattice in its ground state. The low-$T$ value observed in
Fig.~\ref{fig12}(a) is indeed consistent with 
$T=0$ results obtained in a previous
linear spin-wave study.\cite{bul89a} Results for the added-spin model
with $J_{\perp}=J$, shown in Fig.~\ref{fig12}(b), are negative because
of the one extra antiferromagnetic bond, and the size-converged behavior
seems to also tend to a constant as $T\to 0$. This constant value corresponds
to the energy cost of removing the off-plane 
added spin from its host lattice, and its magnitude 
is observed to be roughly one fourth of the low-$T$ value of the vacancy
impurity energy. The $T$ dependence of the $L=64$ results for the
added-spin impurity with $J_{\perp}=J/2$, shown in the inset, are 
qualitatively very similar, but because of the smaller impurity-bond 
strength the absolute values are smaller.

The solid curves in Fig.~\ref{fig12} are results of the corresponding
effective models. Using the same values on $\alpha /J$ and $r/J$ as 
previously when calculating the impurity susceptibilities, we obtain a 
qualitative agreement for the vacancy model while the agreement is 
remarkably good for the added-spin model, both for $J_{\perp}=J$ and 
$J_{\perp}=J/2$ (inset). Hence, in addition to
reproducing the impurity susceptibilities of the full models, the effective
models also describe properly the energetics of the full models. Also, the
parameters $\alpha$ and $r$ can be tuned to give a better agreement for the
vacancy model in Fig.~\ref{fig12}(a), but this in turn will give a poorer
agreement between QMC and effective-model results for the impurity
susceptibility of the vacancy model in Fig.~\ref{fig6}.

\begin{figure}
\includegraphics[width=8cm,clip]{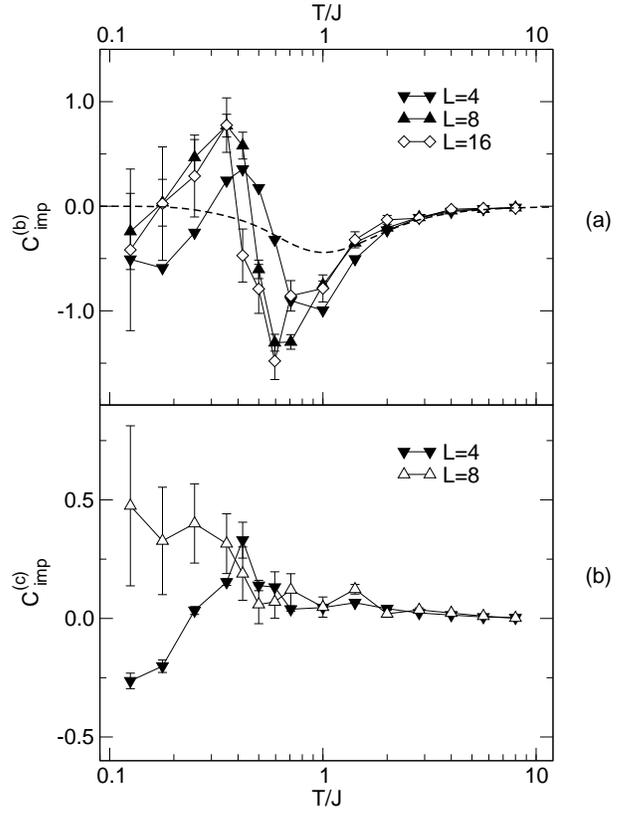}
\caption{The impurity specific heats of the (a) vacancy and (b)
added-spin ($J_{\perp}=J$) models, for different system sizes $L$. The
dashed curve in (a) shows the result of the effective model.}
\label{fig13}
\end{figure}

In Fig.~\ref{fig13} QMC results for the impurity specific heats are shown for
the vacancy model (a) and the added-spin model (b) with $J_{\perp}=J$.
As the system size is increased and the temperature is lowered the statistical
errors grow rapidly. The size-converged behavior is difficult to
determine below $T/J\approx 0.3$, but $C_{\text{imp}}^{(b)}$ in
Fig.~\ref{fig13}(a) is, nevertheless, consistent with the behavior of
$E_{\text{imp}}^{(b)}$ in Fig.~\ref{fig12}(a), as $C=dE(T)/dT$. The 
point at which $C_{\text{imp}}^{(b)}$ goes trough zero, $T/J\approx 0.5$, 
corresponds to the maximum in the energy curve $E_{\text{imp}}^{(b)}$ 
in Fig.~\ref{fig12}. The effective model reproduces well the high-$T$ 
behavior and also exhibits a negative minimum at intermediate temperature.
However, this features is much less pronounced than 
for the full model, and the maximum at lower $T$ is missing. 
In Fig.~\ref{fig12}(b), sufficient accuracy in the simulations has not 
been reached for larger
 system sizes, and the size-converged behavior
can therefore not be determined.

\section{summary}\label{sec_summary}

In this paper we have presented results of an extensive QMC study of impurity
effects in the $S=1/2$ Heisenberg antiferromagnet on a square lattice, as well
as some results for a 3D system. The effects of different types of single
static impurities on the magnetic susceptibility and the specific heat have
been investigated.

For several types of $S_i\neq 0$ impurities in 2D (vacancy, added spin,
ferromagnetically coupled spin), our very precise simulation data has
revealed an additive logarithmic correction to the predicted classical-like
Curie contribution $S_i^2/3T$ to the impurity susceptibility. We have argued
that this logarithmic contribution reflects primarily fluctuations transverse
to the local N\'eel order at the impurity. This is in agreement with recent
field-theoretical work,\cite{sachdev2,sushkov} carried out after our initial
report of log corrections.\cite{paper1} Here we have shown that our numerical
results are in excellent quantitative agreement with these field-theoretical
results,\cite{sachdev2,sushkov} containing both leading and subleading
logarithmic corrections. In 3D, we find no signs of logarithmic corrections,
in accord with predictions.\cite{sachdev2}

In order to have a simple mechanism explaining the leading-order
(i.e., apart from the log corrections) impurity physics, we have also
introduced few-spin effective models. Comparisons with the QMC results
show that the effective models can distinguish between impurities of
different types and spins $S_i$. In many cases the quantitative agreement
between the effective and full models is surprisingly good over a wide
temperature range. This suggests that extended effective models based on
larger clusters of spins, e.g., $3\times 3$ clusters centered around the
site impurities, should give very accurate descriptions, perhaps also for
the vacancy model which we here found was the hardest case to describe
with the simplest effective model.

\acknowledgments{
We are grateful to Subir Sachdev and Oleg Sushkov for very valuable
discussions. This work was supported by the Academy of Finland, project
No.~26175.
}

\appendix*
\section{Selected QMC data for the susceptibility, energy, and specific heat}

\label{sec_app}
\begin{table}
\caption{\label{tab1}Selected $L=64$ data for $\chi _{(i)}^z/L^2$ at
inverse temperature $J/T$, where $i=a,b,$ and $c$ correspond to the
pure, vacancy, and added-spin systems, respectively.}
\begin{ruledtabular}
\begin{tabular}{lllll}
   & & & \multicolumn{2}{c}{$i=c$}\\
  \cline{4-5}
  $J/T$ & $i=a$ & $i=b$  & $J_{\perp}=J$ & $J_{\perp}=J/2$\\
  \hline
  32 & 0.045043(3) & 0.045767(4) & 0.045907(4) & 0.046128(3)\\
  16 & 0.046359(3) & 0.046737(3) & 0.046867(3) & 0.047056(3)\\
  8 & 0.049144(3) & 0.049344(3) & 0.049461(2) & 0.049557(3)\\
  4 & 0.055994(1) & 0.056092(1) & 0.056179(1) & 0.056214(1)\\
  2 & 0.0786001(4) & 0.0786347(4) & 0.0786945(4) & 0.0787106(4)\\
  1 & 0.0935393(2) & 0.0935377(2) & 0.0935859(2) & 0.0935937(2)
\end{tabular}
\end{ruledtabular}
\end{table}

\begin{table}
\caption{\label{tab2}Selected $L=64$ data for $-E_{(i)}/L^2$ at
inverse temperature $J/T$, where $i=a,b,$ and $c$ correspond to the
pure, vacancy, and added-spin systems, respectively.}
\begin{ruledtabular}
\begin{tabular}{lllll}
   & & & \multicolumn{2}{c}{$i=c$}\\
  \cline{4-5}
  $J/T$ & $i=a$ & $i=b$  & $J_{\perp}=J$ & $J_{\perp}=J/2$\\
  \hline
  32 & 0.6694416(5) & 0.6691562(5) & 0.6695154(6) & 0.6694702(5)\\
  16 & 0.6693890(7) & 0.6691048(7) & 0.6694625(7) & 0.6694158(7)\\
  8 & 0.6689102(9) & 0.6686247(8) & 0.668983(1) & 0.6689330(9)\\
  4 & 0.663745(1) & 0.663452(1) & 0.663807(1) & 0.663761(1)\\
  2 & 0.593051(1) & 0.592755(1) & 0.593093(1) & 0.593060(2)\\
  1 & 0.387560(2) & 0.387372(2) & 0.387600(2) & 0.387569(2)
\end{tabular}
\end{ruledtabular}
\end{table}

The numerical data underlying the analyses carried out in this paper are
of very high accuracy---the small errors are only statistical in
nature---and may hence be useful as bench marks for alternative calculations.
In Tables~\ref{tab1} and \ref{tab2} we therefore list $L=64$ data for the
susceptibility and the internal energy, at several inverse temperatures
$J/T$, for the pure (a), vacancy (b), and added-spin models (c).

\begin{figure}
\includegraphics[width=8cm,clip]{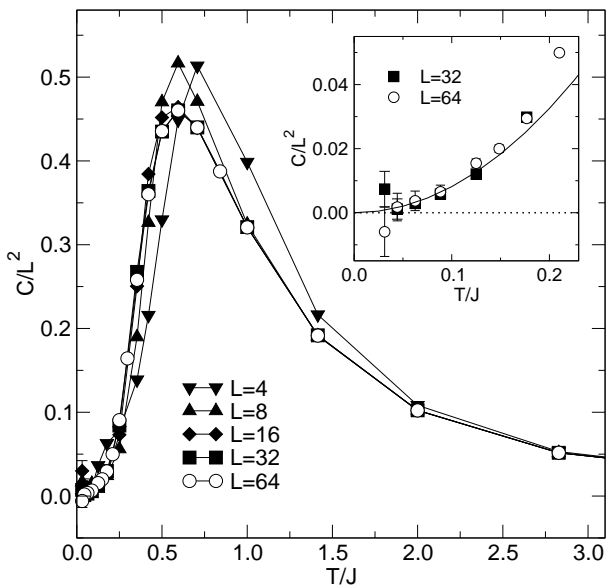}
\caption{Size-normalized specific heats of the pure 2D Heisenberg
antiferromagnet for different system sizes $L$. Error bars are smaller
than the symbols. The inset compares our low-$T$ data with theory
(solid curve)(Ref.~\cite{hasenfratz}).}
\label{fig14}
\end{figure}

In Fig.~\ref{fig14} we show the SSE results for the specific heat of
the pure 2D Heisenberg antiferromagnet at temperatures down to $T/J=1/32$.
At such low temperatures the specific heat has not been determined reliably
in previous studies.\cite{specval} We have obtained the results using
the direct estimator, Eq.~(\ref{spc}). The low-$T$ data shown in the inset of
Fig.~\ref{fig14} are clearly consistent with the quadratic $T$ behavior
suggested in the Hasenfratz-Niedermeyer chiral perturbation theory:
\cite{hasenfratz}
\begin{equation}
C(T)=\frac{6\zeta (3)}{\pi c^2}T^2+O(T^4),
\end{equation}
where we use $c=1.66$ for the spin-wave velocity \cite{velocityvalue} and
$\zeta (3)=1.2020569$.


\begin{thebibliography}{00}

\bibitem{manousakis}
E. Manousakis, Rev. Mod. Phys. \textbf{63}, 1 (1991);
M. A. Kastner, R. J. Birgeneau, G. Shirane, and Y. Endoh,
{\it ibid.} \textbf{70}, 897 (1998).

\bibitem{aharony}
A. Aharony, R. J. Birgeneau, A. Coniglio, M. A. Kastner, and H. E. Stanley, 
Phys. Rev. Lett. \textbf{60}, 1330 (1988).
 
\bibitem{doping}
S.-W. Cheong, A. S. Cooper, L. W. Rupp, B. Batlogg, J. D. Thompson,
and Z. Fisk, Phys. Rev. B {\bf 44}, 9739 (1991);
S. T. Ting, P. Pernambuco-Wise, J. E. Crow, E. Manousakis, and
J. Weaver, {\it ibid.} {\bf 46}, 11772 (1992);
M. Corti, A. Rigamonti, F. Tabak, P. Carretta, F. Licci, and L. Raffo,
{\it ibid.} {\bf 52}, 4226 (1995);
P. Carretta, A. Rigamonti, and R. Sala, {\it ibid.} {\bf 55}, 3734 (1997).

\bibitem{greven}
O. P. Vajk, P. K. Mang, M. Greven, P. M. Gehring, and J. W. Lynn,
Science \textbf{295}, 1691 (2002).

\bibitem{chainexp}
M. Takigawa, N. Motoyama, H. Eisaki, and S. Uchida,
Phys. Rev. B {\bf 55}, 14129 (1997).

\bibitem{ladderexp}
M. Azuma, Y. Fujishiro, M. Takano, M. Nohara, and H. Takagi,
Phys. Rev. B {\bf 55}, R8658 (1997).

\bibitem{castro_neto}
A. H. Castro Neto, E. Novais, L. Borda, G. Zar\'{a}nd, and I. Affleck,
Phys. Rev. Lett. \textbf{91}, 096401 (2003).

\bibitem{egg92a}
S. Eggert and I. Affleck, Phys. Rev. B {\bf 46}, 10866 (1992).

\bibitem{iga95a}
J. Igarashi, T. Tonegawa, M. Kaburagi, and P. Fulde,
Phys. Rev. B {\bf 51}, 5814 (1995).

\bibitem{nis00a}
M. Nishino, H. Onishi, P. Roos, K. Yamaguchi, and S. Miyashita, 
Phys. Rev. B {\bf 61}, 4033 (2000).

\bibitem{nis00b}
M. Nishino, H. Onishi, K. Yamaguchi, and S. Miyashita, 
Phys. Rev. B {\bf 62}, 9463 (2000).

\bibitem{fuk96a}
H. Fukuyama, N. Nagaosa, M. Saito, and T. Tanimoto,
J. Phys. Soc. Jpn. {\bf 65}, 2377 (1996).

\bibitem{san97a}
A. W. Sandvik, E. Dagotto, and D. J. Scalapino, Phys. Rev. B 
{\bf 56}, 11701 (1997).

\bibitem{mar97a}
G. B. Martins, M. Laukamp, J. Riera, and E. Dagotto, 
Phys. Rev. Lett. \textbf{78}, 3563 (1997).

\bibitem{bul89a}
N. Bulut, D. Hone and D. J. Scalapino, and E. Y. Loh, Phys. Rev. Lett.
\textbf{62}, 2192 (1989).

\bibitem{nagaosa}
N. Nagaosa, Y. Hatsugai, and M. Imada, J. Phys. Soc. Jpn. \textbf{58},
978 (1989).

\bibitem{sch94a}
P. Schlottmann, J. Appl. Phys. {\bf 75}, 5532 (1994).

\bibitem{kot98a}
V. N. Kotov, J. Oitmaa, and O. Sushkov, Phys. Rev. B {\bf 58}, 8495 (1998);
{\it ibid}. {\bf 58}, 8500 (1998).

\bibitem{nag95a}
N. Nagaosa and T.-K. Ng, Phys. Rev. B \textbf{51}, 15588 (1995).

\bibitem{sus00a}
O. P. Sushkov, Phys. Rev. B {\bf 62}, 12135 (2000).

\bibitem{sachdev1}
S. Sachdev, C. Buragohain, and M. Vojta, Science \textbf{286},
2479 (1999); M. Vojta, C. Buragohain, and S. Sachdev, Phys. Rev.
B \textbf{61}, 15152 (2000).

\bibitem{sac01a}
S. Sachdev, M. Troyer, and M. Vojta, Phys. Rev. Lett. \textbf{86}, 
2617 (2001).

\bibitem{egg95a}
S. Eggert and I. Affleck, Phys. Rev. Lett. {\bf 75}, 934 (1995).

\bibitem{egg98a}
S. Eggert and S. Rommer,  Phys. Rev. Lett. {\bf 81}, 1690 (1998);
S. Rommer and S. Eggert, Phys. Rev. B {\bf 59}, 6301 (1999).

\bibitem{fuj03a}
S. Fujimoto and S. Eggert, Phys. Rev. Lett. {\bf 92}, 037206 (2004).

\bibitem{mur97a}
K. Murayama and J. Igarashi, J. Phys. Soc. Jpn. {\bf 66}, 1157 (1996).

\bibitem{paper1}
K. H. H\"oglund and A. W. Sandvik, Phys. Rev. Lett. \textbf{91},
077204 (2003).

\bibitem{sachdev2}
S. Sachdev and M. Vojta, Phys. Rev. B \textbf{68}, 064419 (2003).

\bibitem{sushkov}
O. P. Sushkov, Phys. Rev. B \textbf{68}, 094426 (2003).

\bibitem{sepnote1}
A separation  of the impurity susceptibility in components parallel
and perpendicular to the N\'eel order is of course strictly applicable
only at $T=0$, i.e., when the spin-rotation symmetry is broken. It is
nevertheless a useful concept also when the impurity is coupled to a
large ordered domain with slow dynamics. However, the exact way in which
the separation into longitudinal and transverse components is done is to
some extent a matter of calculational definitions outside the limit $T=0$.

\bibitem{harris}
A. B. Harris and S. Kirkpatrick, Phys. Rev. B {\bf 16}, 542 (1977).

\bibitem{chernyshev}
A. L. Chernyshev, Y. C. Chen, and A. H. Castro Neto,
Phys. Rev. B \textbf{65}, 104407 (2002).

\bibitem{sse1}
A. W. Sandvik, Phys. Rev. B \textbf{56}, 11678 (1997).

\bibitem{sse2}
A. W. Sandvik, Phys. Rev. B \textbf{59}, R14157 (1999).

\bibitem{sandvik3d}
A. W. Sandvik, Phys. Rev. Lett. \textbf{80}, 5196 (1998).

\bibitem{chn}
S. Chakravarty, B. I. Halperin, and D. R. Nelson, Phys. Rev. Lett.
\textbf{60}, 1057 (1988).

\bibitem{sepnote2}
We here use slightly different values for $\alpha$ and $r$ than
previously in Ref.~\onlinecite{paper1}, in order to optimize the agreement
between QMC and effective model results for both the susceptibility
and the energy, as well as to get a good agreement also in the case of
the added-spin model with $J_{\perp}=J/2$. The results for the vacancy
and the added spin with $J_{\perp}=J$ change very little from those in
Ref.~\onlinecite{paper1}.

\bibitem{sse3}
P. Henelius and A. W. Sandvik, Phys. Rev. B \textbf{62}, 1102 (2000).

\bibitem{sse4}
A. W. Sandvik, Phys. Rev. B \textbf{50}, 15803 (1994).

\bibitem{evertz}
H. G. Evertz, Adv. Phys. \textbf{52}, 1 (2003).

\bibitem{hasenfratz}
P. Hasenfratz and F. Niedermayer, Z. Phys. B \textbf{92}, 91 (1993).

\bibitem{stiffnessvalue}
A. W. Sandvik, Phys. Rev. B {\bf 66}, 024418 (2002).

\bibitem{specval}
J.-K. Kim and M. Troyer, Phys. Rev. Lett. \textbf{80}, 2705 (1998);
J. Jakli\v{c} and P. Prelov\v{s}ek, {\it ibid.} \textbf{77}, 892 (1996);
G. Gomez-Santos, J. D. Joannopoulos, and J. W. Negele, Phys. Rev. B
\textbf{39}, 4435 (1989).

\bibitem{velocityvalue}
An accurate value of the spin-wave velocity $c=\sqrt{\rho_s/\chi_\perp}$
is obtained from the stiffness $\rho_s \approx 0.181$ calculated in
Ref.~\onlinecite{stiffnessvalue} and the perpendicular susceptibility
$\chi_{\perp}\approx 0.0659$ obtained by O. F. Sylju{\aa}sen and
A. W. Sandvik, Phys. Rev. E {\bf 66}, 046701 (2002).

\end{thebibliography}
\end{document}